\documentclass[journal]{IEEEtran}

\usepackage[usenames, dvipsnames]{xcolor} %plus de couleurs...
\usepackage{tikz}
\usetikzlibrary{calc}
\usetikzlibrary{plotmarks}
\usetikzlibrary{shapes,arrows}

\def\SizeMark{2}
\def\SizePlot{2}
\usepackage[cmex10]{amsmath}
\usepackage{amsfonts}
\usepackage{dsfont}
\newtheorem{prop}{Proposition}
\newtheorem{assump}{Assumption}

\usepackage{graphicx}
\usepackage[noadjust]{cite}
\usepackage{bm}
\usepackage{bbm}
\usepackage{yhmath}
\usepackage{amssymb}
\usepackage{amscd}

\begin{document}
\title{Charging Games in Networks of Electrical Vehicles}

\author{Olivier~Beaude, Samson~Lasaulce, and Martin Hennebel~
\thanks{O. Beaude is with L2S, Renault, and Supelec, France~; S. Lasaulce is with CNRS, L2S, France~; M. Hennebel is with Supelec, France. E-mail @~: olivier.beaude@lss.supelec.fr, lasaulce@lss.supelec.fr, martin.hennebel@supelec.fr}% <-this % stops a space
%\thanks{}% <-this % stops a space
}
%\thanks{Manuscript received October 8, 2012}}

\maketitle

\begin{abstract}

In this paper, a static non-cooperative game formulation of the problem of distributed
charging in electrical vehicle (EV) networks is proposed. This formulation allows one to model the interaction between several EV which are connected to a common residential distribution transformer. Each EV aims at choosing the time at which it starts charging its battery in order to minimize an individual cost which is mainly related to the total power delivered by the transformer, the location of the time interval over which the charging operation is performed, and the charging duration needed for the considered EV to have its battery fully recharged. As individual cost functions are assumed to be memoryless, it is possible to show that the game of interest is always an ordinal potential game. More precisely, both an atomic and nonatomic versions of the charging game are considered. In both cases, equilibrium analysis is conducted. In particular, important issues such as equilibrium uniqueness and efficiency are tackled. Interestingly, both analytical and numerical results show that the efficiency loss due to decentralization (e.g., when cost functions such as distribution network Joule losses or life of residential distribution transformers when no thermal inertia is assumed) induced by charging is small and the corresponding "efficiency", a notion close to the Price of Anarchy, tends to one when the number of EV increases.

\emph{Keywords: Charging games, electrical vehicle, distribution networks, potential games, Nash equilibrium, price of anarchy.}
\end{abstract}

%%%%%%%%%%%%%%%%%%%%%%%%%%%%%%%%%%%
\section{Introduction}
\label{sec:introduction}
%%%%%%%%%%%%%%%%%%%%%%%%%%%%%%%%%%%

For various reasons which include the decrease of fossil fuel production, the design of electric vehicles (EV) and plug-in hybrid electric vehicles (PHEV) becomes a more and more important issue. An intrinsic feature of EV is that they need their battery to be recharged regularly. A critical issue is that the charging power is comparable to the
 maximum power corresponding to a typical consumer's subscription. In France, for instance, the former is typically about 3 kW at home while the latter is about 6 kW (or kVA to be more precise). This shows the importance of scheduling in an appropriate manner the charging period \cite{Clement-Nyns2009}. This is precisely what this paper is about. As the decision to plug the EV to the network and to start charging belongs to the vehicle owner, the problem is naturally distributed. This is one of the reasons why a reasonable mathematical model to analyze such a problem is given by non-cooperative games. It is quite recent that game-theoretic tools have been applied to smart grids (see e.g., \cite{Saad2012} for a recent survey). Interesting contributions include \cite{Wu2012,Wu2011,Agarwal2011,Mohsenian-Rad2010,Vytelingum2010}. As far as the problem of charging is concerned, the authors of \cite{Wu2011} show the usefulness of a well-chosen pricing policy to incite users to charge their vehicle in order to regulate frequency of the distribution network. References \cite{Agarwal2011}\cite{Mohsenian-Rad2010} used a similar method to study the more general problem of load balancing.

Compared to the application-oriented works where game theory is used to optimize energy consumption at the user side (at home, by the EV, etc), the present work has the following features. First, contrary to the vast majority of game-theoretical studies in this area which apply to generic energy consumption problems, specific features of the distributed charging problem are accounted for~: to avoid premature aging of the EV battery and make the EV available, the EV battery has to be charged without interruption (the charging profile is imposed to be a rectangular function) and within a given time windows. Second, to the best of our knowledge, there seems to be no work available in the literature where distributed charging policies are optimized in terms of the considered distribution network physical costs with the use of non-cooperative games. These costs are assumed to be memoryless in the sense that at a given time they do not depend on the past sequence of distribution network load levels. These costs include the life of the distribution transformer to which the vehicles are (indirectly) connected and Joule losses in the transformer and in the distribution lines between the transformer and the charging points. 

Mathematically, the game under investigation has a structure which is close to a congestion game \cite{Rosenthal1973}. Indeed, time instances at which an EV can charge its battery can be seen as a set of available facilities "simultaneously". However, since several facilities can be exploited, it is not a congestion game in the sense of Rosenthal \cite{Rosenthal1973}. The closest model is rather given by \cite{Orda1993}. The latter is concerned with the more general scenario in which each user can split its demand between several facilities (called "parallel links" in \cite{Orda1993}). The main technical differences between the game under investigation and the latter is that the set of facilities has to be contiguous and less symmetry is available which makes both the cost/utility functions and constraints different. One of the consequences of this is that a notion of "efficiency" close to the price of anarchy \cite{Koutsoupias1999} of the game considered in this paper cannot be upper bounded by using classical results such as \cite{correa-mor-2004}.

In this paper, two versions of the considered charging game will be studied, namely~: an atomic version which assumes a finite number of EV~; a nonatomic version where only fractions of users (those number is therefore implicitly assumed to be large) appear. The nonatomic version is both useful to analyze dense residential networks but also to obtain insights onto the finite game e.g., on how the price of anarchy behaves. Sec. \ref{sec:model} and Sec. \ref{sec:atomic-charging-game} correspond to the case of a finite number of EV while Sec. \ref{sec:nonatomic-charging-game} is dedicated to the limiting case of the nonatomic scenario.

\emph{Notations~:} bold symbols $\mathbf{X},\mathbf{x}$ will stand for vectors while the notation $x_t$ will refer to the $t-$th component of $\mathbf{x}$.

%%%%%%%%%%%%%%%%%%%%%%%%%%%%%%%%%%%
\section{Proposed system model}
\label{sec:model}
%%%%%%%%%%%%%%%%%%%%%%%%%%%%%%%%%%%

Consider a residential distribution transformer which has to deliver electrical power to a set of households and EV. The current or power delivered by the transformer is assumed to have two components~: on component due to the set of EV and another one which is due to the other electrical equipments. The latter will be referred to as an exogenous  component because it is independent of the EV charging policies. Time is assumed to be slotted (a time-slot duration is typically 30 min). The time index is denoted by $t$ and belongs to
\begin{equation}
\label{DescrT}
\mathcal{T}=\left\{1,2,...,T\right\}.
\end{equation}
For example, if the time windows under consideration is from 5 pm (day number $j$) to 8 am (day number $j+1$), there are $T=31$ time instances or 30 min time-slots at which an EV may be active or not.

\noindent

In the atomic formulation of the charging problem the set of EV will be denoted by
\begin{equation}
\mathcal{I}^{\text{A}} =\left\{1, 2, ..., I\right\}.
\end{equation}

The arrival and departure time of EV $i \in \mathcal{I}^{\text{A}}$ will be denoted by $a_i \in \mathcal{T}$ and $d_i \in \mathcal{T}$ respectively. As it is assumed that an EV has to charge its battery within the total time windows, EV $i$ chooses the time instance at which it starts charging in the following action set~:
\begin{equation}
\label{DebCharge}
s_{i} \in \mathcal{S}_{i}=\left\{a_{i},a_{i}+1,...d_{i}-C_{i}\right\}
\end{equation}
where $C_i$ represents the number of time instances or time-slots required to have the battery charged or reach a required state of charge (SoC) for the next trip of EV $i$. A special case is when all users have the same charging constraint, that is $\forall i, \, a_{i}=a, d_{i}=d, C_{i}=C$. This case will be said to be symmetric. In this case, it will be assumed, without loss of generality, that $a=1$ and $d=T$.

\noindent

For simplicity, the charging power is assumed to have two possible levels namely $0$ or $P$. As already mentioned, charging profiles are assumed to be rectangular. One technological reason for this is to manage the EV battery life which is accelerated when time-varying charging power levels are allowed. Summing up, a user decides when to plug his vehicle by choosing $(a_i, d_i) \in \mathcal{T}^2$ and this defines the action or strategy set for the EV $i$. The action or strategy profile is denoted by
\begin{equation}
\bm{s}=\left(s_1, s_2, ..., s_I \right)
\end{equation}
which lies in
\begin{equation}
\mathcal{S}=\prod_{i=1}^{I}  \mathcal{S}_{i}.
\end{equation}
The standard notation $\bm{s_{-i}}=\left(s_{1},s_{2},...,s_{i-1},s_{i+1},..,s_{I}\right)$ for referring to the action profile in which user $i$'s action is removed will be used.

Given EV charging decisions, let respectively $\tilde{n}_{t}, \, n_{t}$ denote the numbers of EV starting to charge and charging at time $t$. In the atomic case, these quantities are related to $s_{i}$ by
\begin{equation}
\label{NEVBeginningCharge}
\tilde{n}_{t}(\bm{s})=\displaystyle \sum_{i=1}^{I} \mathds{1}_{s_{i}=t} \textrm{ ,}
\end{equation}
and
\begin{equation}
\label{NEVCharging}
n_{t}(\bm{s}) = \displaystyle \sum_{i=1}^{I} \displaystyle \sum_{t'=1}^{C_{i}} \mathds{1}_{{s_{i}=t-C_{i}+t'}}
\end{equation}
where $\mathds{1}_{s_{i}=t}$ is the indicator function.

\noindent

To evaluate the impact of the charging policies on the distribution network, the key quantity is the total transformer load or consumed power. The $T-$dimensional sequence of load levels $\bm{L}(\bm{s})= \left(L_1(\bm{s}), L_2(\bm{s}), ..., L_T(\bm{s}) \right)$ corresponding to the period of time under consideration expresses as
\begin{equation}
\label{TotLoad}
\bm{L}(\bm{s})=\bm{L^{\text{exo}}}+P\bm{n}(\bm{s})
\end{equation}
where $\bm{L^{\text{exo}}}$ is the sequence of exogenous loads which are not due to the presence of the EV and $\bm{n}(\bm{s}) $ is the sequence of the numbers of active or charging vehicles. It is implicitly assumed that the location of the exogenous loads and EV do not need to be accounted for according to the above model, which is realistic e.g., when the considered network cost is given by Joule losses in networks with a symmetric topology or if the main physical cost is due to transformer losses (such as copper losses, iron losses, or transformer life). Under this assumption, the total impact on the grid on the considered period is then 
\begin{equation}
\label{CostGrid}
\textrm{TC}^{\textrm{Grid}}(\bm{s})=\displaystyle \sum_{t \in \mathcal{T}} f^{\textrm{Grid}}(L_{t}(\bm{s})) \textrm{.}
\end{equation}

%%%%%%%%%%%%%%%%%%%%%%%%%
\section{Atomic charging game~: proposed formulation and main results}
\label{sec:atomic-charging-game}
%%%%%%%%%%%%%%%%%%%%%%%%%

%---------------------------------
\subsection{Game description}
\label{subsec:formulation}

The problem described in the preceding section is modeled by a static non-cooperative game under strategic form (see e.g., \cite{Lasaulce-Tutorial-09}\cite{Lasaulce-Tembine-Book-2011}). The game can therefore be described mathematically by a triplet $\mathcal{G}^{\text{A}}=(\mathcal{I}, \{ \mathcal{S}_i \}_{i\in\mathcal{I}}  ,\{ u_i \}_{i\in\mathcal{I}} )$ whose components are defined as follows~:

\begin{itemize}
  \item the set of players is given by the set of electrical vehicles $\mathcal{I}$~;
  \item the action/strategy set for player $i\in \mathcal{I}$ is defined by (\ref{DebCharge})~;
  \item the utility function for player $i \in \mathcal{I}$ is defined by
\begin{equation}
\label{PersonalCostC}
u_{i}(s_{i},s_{-i})=-g_{i}\left(\displaystyle \sum_{t=s_{i}}^{s_{i}+C_{i}-1}f^{\text{Grid}}(L_{t}(\bm{s}))\right)
\end{equation}  
\end{itemize}
where $f^{\text{Grid}}$ translates the vehicle needs in terms of load or consumed power into a physical cost for the grid and more precisely for the residential distribution network. For typical physical costs such as losses in the distribution transformer and over the distribution lines or the transformer life, the following assumptions are valid.

\medskip

\begin{assump}
\label{H1}
$f^{\text{Grid}}$ is continuous and strictly increasing.
\end{assump}

\medskip

\begin{assump}
\label{H2}
$f^{\text{Grid}}$ is continuously differentiable and strictly convex.
\end{assump}

\medskip

As for the function $g_i$, it corresponds to a perceived cost from the user standpoint~: it may be a monetary conversion or pricing function. It is assumed to be strictly increasing. As a last comment, note that the proposed utility function means that the user is charged with a cost which is directly related to the time period he chooses to charge the battery of his EV, and do not depend on the total load during other time periods.

%---------------------------------------
\subsection{Equilibrium analysis}
\label{subsec:equilibrium-analysis}

The purpose of this section is to analyze important issues such as existence, uniqueness, and efficiency of Nash equilibrium (the reader could refer to \cite{Lasaulce-Tembine-Book-2011} for a reminder of this concept). Remarkably, although the game under investigation is not a congestion game, it is always an ordinal potential game \cite{Monderer1996} and can also be an exact potential game in some special cases. It thus inherit many favorable properties described in the seminal work of \cite{Monderer1996}. Let us state the first Proposition.

\medskip

\begin{prop}
\label{PropPotGame}
The charging game $\mathcal{G}^{\text{A}}$ is an ordinal potential game for which a possible potential function is given by
\begin{equation}
\label{PotentialAt}
\Phi^{\textrm{A}}\left(\bm{s}\right) = -\displaystyle \sum_{t \in \mathcal{T}} \displaystyle \sum_{v_{t}=0}^{n_{t}(\bm{s})}f^{\text{Grid}}\left(L_{t}^{\text{exo}}+Pv_{t}\right) \textrm{,}
\end{equation}
\end{prop}

\medskip

To prove this result, the fact that $g_i$ is a strictly increasing function is exploited. It can be effectively verified that a deviation of user $i$'s charging strategy $s_{i} \rightarrow s^{\prime}_{i}$ translates into a variation
\begin{equation}
u_{i}(s_{i},s_{-i})-u_{i}(s^{\prime}_{i},s_{-i}) \textrm{,}
\end{equation}
whose sign is identical to
\begin{equation}
\Phi^{\textrm{A}}(s_{i},s_{-i})-\Phi^{\textrm{A}}(s^{\prime}_{i},s_{-i}) \textrm{.}
\end{equation}

Note that if the pricing function $g_i$ is assumed to be the identity function then $\mathcal{G}^{\text{A}}$ is also an exact potential game. Since the atomic charging game is always an ordinal potential game, the existence of a pure Nash equilibrium is guaranteed. Indeed, from \cite{Monderer1996} the next proposition follows.

\medskip

\begin{prop}
\label{PropExistPureNE}
The charging game $\mathcal{G}^{\text{A}}$ has always a pure Nash Equilibrium.
\end{prop}

\medskip

The set of Nash equilibria of this game will be denoted by $\mathcal{S_{\textrm{E}}}$. An important issue, especially when converging iterative charging algorithms have to be designed, is to know whether this set is a singleton. Note that in the following and similarly to the notion of uniqueness used in routing games for example, when we will refer to uniqueness it will be for the resulting charging configuration $\bm{n}$ and not for the strategic profile $\bm{s}$\footnote{Uniqueness in terms of the strategy profile $\bm{s}$ is stronger and implies uniqueness in terms of charging configuration $\bm{n}$. Because of the symmetry between players, it may indeed exist various strategy profiles $\bm{s}$ providing a same configuration $\bm{n}$ and this would avoid obtaining uniqueness results when considering the notion related to $\bm{s}$.}. With this definition, it turns out that uniqueness does not hold in general, especially when the number of EV is small. Here is a simple counter-example which sustains this assertion.

\emph{Counter-example for uniqueness.} Consider the case where $f^{\text{Grid}}$ has to account for Joule losses~:
\begin{equation}
f^{\text{Grid}}(L_{t})=R L_{t}^2 \textrm{.}
\end{equation}
Assume that $T=6$, $\bm{L^{exo}}=(1,2,3,2,1,3)$, $I=3$, $\forall i$, $C_{i}=2$, $a_{i}=0$, $d_{i}=6$, and $P=1$. With these parameters, the following two sequences of number of active EV 
\begin{equation}
\bm{n}^{1}=(3;4;3;3;2;0) \text{ and } \bm{n}^{2}=(2;3;3;4;3;0)
\end{equation}
can be checked to correspond to Nash Equilibria. 
  
\noindent  

Since uniqueness is not guaranteed in general, studying the number of Nash equilibria and their efficiency -through a notion close to the price of anarchy- is a relevant issue. However, this type of analysis is not trivial. Therefore, to obtain some insights into the problem under investigation we will analyze the limiting case where the number of EV becomes large. Additionally, a detailed numerical study will be provided to better understand both the atomic and nonatomic cases.

%%%%%%%%%%%%%%%%%%%%%%%%%
\section{The nonatomic charging game}
\label{sec:nonatomic-charging-game}
%%%%%%%%%%%%%%%%%%%%%%%%%

As explained in the preceding sections, our motivations for studying the case where the number of EV is large is twofold. First, it may correspond to real scenarios where the distribution network is dense. Second, the nonatomic case provides more insights on the atomic case. 

%%%%%%%%%%%%%%%%%%%%%%%%%
\subsection{Description of the nonatomic charging game}
%%%%%%%%%%%%%%%%%%%%%%%%%

To define this new game, let us introduce the following quantities.

In the nonatomic formulation, a continuum of EV is considered and will be denoted by
\begin{equation}
\mathcal{I}^{\text{NA}} =\left[0;1\right] \text{,}
\end{equation} 
with a set $\mathcal{K}$ of classes of users with respective demands $w_{k}$. These demands sum to $1$ given that $\mathcal{I}^{\text{NA}}=\left[0;1\right]$. The arrival and departure time $a_{k}$ and $d_{k}$, as well as the charging requirement $C_{k}$ will now be related to a class of user $k$, while it was related to a single user in the atomic framework.

The action set of each class $\mathcal{S}_{k}$ is defined as in (\ref{DebCharge}). Let $x_{kt}$ define the proportion of EV of class $k$ choosing time $t$ to begin charging. $\bm{x}_{k}=(x_{kt})_{t \in \mathcal{T}}$ verifies
\begin{equation}
\forall k \in \mathcal{K}, \, \bm{x}_{k} \in \left[0;1\right]^{T} \text{ with } 
\left\{
\begin{array}{l}
\forall t \notin \mathcal{S}_{k}, \, x_{kt}=0 \\
\displaystyle \sum_{t=a_{k}}^{d_{k}-C_{k}} x_{kt} = 1 \\
\end{array} 
\right.
\text{,}
\end{equation} 
which means that $\bm{x_{k}}$ belongs to the $(d_{k}-C_{k}-a_{k}+1)$-dimensional simplex. 

The proportion of EV starting to charge and charging at time $t$, denoted respectively by $\tilde{x}_t(\bm{s})$ and $x_t(\bm{s})$, follow then directly with 
\begin{equation}
\label{NEVBeginningChargeNonAt}
\tilde{x}_t(\bm{s})=\displaystyle \sum_{k \in \mathcal{K}} w_{k}s_{kt} \textrm{ ,}
\end{equation}
and
\begin{equation}
\label{NEVChargingNonAt}
x_{t}(\bm{s}) = \displaystyle \sum_{k \in \mathcal{K}}w_{k} \displaystyle \sum_{t'=1}^{C_{k}}s_{k(t-C_{k}+t')}
\end{equation}

Note that when the symmetric case is considered, $\bm{\tilde{x}}=(\tilde{x}_{t})_{t \in \mathcal{T}}$ lies in the $(T-C)$-dimensional simplex, denoted by $\Delta_{T-C}$\footnote{This is not the case for $\bm{x}$ because its first and last $C$ components are ordered given the charging constraint.}.

\medskip

The total transformer load is then expressed as
\begin{equation}
\label{TotLoadNonAt}
\bm{L}(\bm{x})=\bm{L^{\text{exo}}}+P\bm{x}
\end{equation}

and the utility function for an EV $i$ of class $k$ when choosing to start charging at $t=s_{i}$ is

\begin{equation}
\label{PersonalCostCNonAt}
u_{k}(s_{i},\bm{x})=-g_{i}\left(\displaystyle \sum_{t=s_{i}}^{s_{i}+C_{k}-1}f^{\text{Grid}}(L_{t}(\bm{x}))\right)
\end{equation} 

Note that here, notation $s_{-i}$ has no more sense : because action of EV $i$ has a negligible effect on the total load, $s_{-i}$ is in fact the whole charging decision vector $\bm{s}$.

\noindent

In this setting, a Nash equilibrium is a configuration $\bm{x}^{\text{NE}}$ for which each strategy $t$ played with a positive weight, that is $\displaystyle \sum_{k \in \mathcal{K}} x^{\text{NE}}_{kt}>0$, has a minimum cost :
\begin{equation}
\forall t, \, \forall k \in \mathcal{K} : \: x^{\text{NE}}_{kt}>0, t \in \arg\min_{t^{\prime} \in \mathcal{S}_{k}} u_{k}(t^{\prime},\bm{x}^{\text{NE}})
\end{equation}

%%%%%%%%%%%%%%%%%%%%%%%%%
\subsection{Equilibrium analysis of the nonatomic charging game}
%%%%%%%%%%%%%%%%%%%%%%%%%

The main properties of the nonatomic charging game are now presented.

\medskip

\begin{prop}\textit{(Potential property of the nonatomic charging game)}
\label{PropPotGameNonAt}
The nonatomic charging game is also an ordinal potential game for which a possible potential function is

\begin{equation}
\label{PotentialNonAt}
\Phi^{\textrm{NA}}\left(\bm{s}\right) = -\displaystyle \sum_{t \in \mathcal{T}} \displaystyle \int_{v_{t}=0}^{x_{t}(\bm{s})}f^{Grid}\left(L_{t}^{exo}+Pv_{t}\right)dv_{t} \textrm{.}
\end{equation}
\end{prop}

\medskip

Similarly to the atomic case, the potential property ensures then that the nonatomic charging game has a pure Nash equilibrium. 

\medskip

\begin{prop}\textit{(Existence of a pure nonatomic Nash Equilibrium)}
\label{PropExistPureNENonAt}
The nonatomic charging game has a pure Nash Equilibrium.
\end{prop}

\medskip

Moreover, and on the contrary to the atomic setting, uniqueness is also valid under the weak Ass.\ref{H1}.

\medskip

\begin{prop}\textit{(Uniqueness of nonatomic Nash Equilibrium)}
\label{PropUniquNonAtNE}
When $f^{Grid}$ verifies Ass.\ref{H1}, the nonatomic charging game has a unique Nash Equilibrium. 
\end{prop}

\medskip

\begin{IEEEproof}
Given that $\bm{x}$ belongs to a convex set and that $\Phi^{\textrm{NA}}$ is strictly concave under Ass.\ref{H1}, $\Phi^{\textrm{NA}}$ admits a unique maximum. The uniqueness of the Nash equilibrium comes then straightforward because Nash equilibria of the game are in the set of local maxima of $\Phi$, due to the potential property (\ref{PropPotGameNonAt}).
\end{IEEEproof}

\medskip

In fact, when considering a symmetric continuum of players, that is parameters $C, \, \alpha, \, \beta$ are the same for all classes of EV, uniqueness even holds in the class of functions satisfying Ass.\ref{H1} : whatever the function in this class, the unique nonatomic Nash equilibrium of the charging game will be the same. This nevertheless needs to introduce an additional hypothesis on the exogeneous load. To this end, the Euclidean division of $T-C+1$ by $C$ is now defined by

\begin{equation}
T-C+1=q_{T,C}C+r_{T,C} \textrm{ with } r_{T,C}<C \textrm{,}
\end{equation}

\medskip

\begin{prop}\textit{(Uniqueness of nonatomic Nash Equilibrium in the class of functions (\ref{H1}))}
\label{PropUniquNENonAtClass1}
In the symmetric nonatomic case, suppose that the following conditions are valid
\begin{itemize}
\item $\bm{L^{exo}}$ is convex and increasing,
\item $\bm{L^{exo}}$ is such that
\begin{equation}
\label{CondSufI}
1>q_{T,C}L_{T-1}^{exo}-\displaystyle \sum_{k=1}^{q_{T,C}}L_{T-1-kC}^{exo} \textrm{.}
\end{equation}
\end{itemize}
then the Nash equilibrium is the same for all functions verifying Ass.\ref{H1}. 
\end{prop}

\medskip

The proof is given in Appendix\ref{ProofUniqNEClass1}. 

\medskip

Observe that this result applies for example in the case of a constant $\bm{L^{exo}}$, Equation(\ref{CondSufI}) becoming $1>0$. Naturally, this is no more true for more general profiles of $\bm{L^{exo}}$, as presented in the following example with $T=11$, $C=5$
\begin{equation}
\bm{L^{exo}}=(0.1,0.2,0.3,0.4,0.5,0.2,0.2,0.3,0.2,0.1,0.2) \text{.}
\end{equation}

Consider
\begin{equation}
f_{1}^{Grid}(L_{t})=\sqrt{L_{t}}, f_{2}^{Grid}=(L_{t})^8 \textrm{,}
\end{equation}

both verifying Ass.\ref{H1}, but leading to the two different nonatomic Nash equilibria

\begin{equation}
\bm{\tilde{x}_{1}}^{\textrm{NE}}=(0.45,0,0,0,0,0.55,0,0,0,0,0) \textrm{ and } 
\end{equation}

\begin{equation}
\bm{\tilde{x}_{2}}^{\textrm{NE}} = (0.42,0,0,0,0,0.58,0,0,0,0,0)
\end{equation}

%%%%%%%%%%%%%%%%%%%%%%%%%
\subsection{"Efficiency" equals one in the symmetric nonatomic charging game}
%%%%%%%%%%%%%%%%%%%%%%%%%

The \textit{Price of Anarchy} (PoA) \cite{Koutsoupias1999} measures how the efficiency of a system degrades due to selfish behavior of its agents. This is very important to decide whether a decentralized mechanism can be applied, regarding the loss of efficiency in comparison with the performance that would be obtained with a central authority. In the problem considered here, this notion will be slightly different and defined as the "efficiency" of the game

\begin{equation}
\label{PoA}
\textrm{Efficiency}=\frac{\max_{\bm{s} \in \mathcal{S_{\textrm{E}}}}\textrm{TC}^{\textrm{Grid}}(\bm{s})}{\min_{\bm{s}\in \mathcal{S}}\textrm{TC}^{\textrm{Grid}}(\bm{s})} \textrm{,}
\end{equation} 

with $\textrm{TC}^{\textrm{Grid}}$ given by (\ref{CostGrid}).

\medskip

With this metric, we search for a bound on the efficiency. Observe that contrary to the standard literature in this field, variational inequalities methods do not apply directly, because the social objective is not weighted by the load on respective time slots $L_{t}$. The following result shows that if $f^{Grid}$ is in the class of functions verifying Ass.\ref{H1}-\ref{H2}, the symmetric nonatomic efficiency is one.

\medskip

\begin{prop}\textit{(Symmetric nonatomic efficiency equals one)}
\label{PropPoA}
In the nonatomic symmetric case, with $f^{Grid}$ verifying Ass.\ref{H1}-\ref{H2} and $\bm{L^{exo}}$ verifying the same conditions than in Prop.\ref{PropUniquNENonAtClass1}, the efficiency of the charging game is one.
\end{prop}

\medskip

\begin{IEEEproof}
Due to the potential property, the Nash equilibrium of the game $\mathcal{G}=(\mathcal{I},\mathcal{S},(f^{Grid})^{\prime})$ is the social optimum when considering costs $f^{Grid}$. Because $f^{Grid}$ and $(f^{Grid})^{\prime}$ both verify Ass.\ref{H1} when Ass.\ref{H1}-\ref{H2} hold for $f^{Grid}$, Prop.\ref{PropUniquNENonAtClass1} applies and their Nash equilibrium concide which concludes the proof. 
\end{IEEEproof}

\medskip

Tight bounds have been determined for the standard notion of Price of Anarchy in congestion games\cite{Nisan2007}, with the standard definition based on total cost weighted by the number of users choosing each alternative. For example, in the case of quadratic cost functions with nonnegative coefficients, the provided value is $\frac{3\sqrt{3}}{3\sqrt{3}-2} \approx 1.626$ and this bound increases with the maximal power of the polynomial costs. In this setting, the condition imposing that a charging vector must not be fractioned, which consists in choosing $C$ contiguous alternatives, provides this better bound of one.

%%%%%%%%%%%%%%%%%%%%%%%%%
\section{Numerical analysis}
\label{sec:numerical-analysis}
%%%%%%%%%%%%%%%%%%%%%%%%%

To investigate in more details the properties of the atomic charging game, for which an analytical value of the efficiency has not been proven to exist, a numerical analysis has been conducted. To easily compare the results obtained in these simulations with the analytical ones of the nonatomic case, $\bm{L^{exo}}=0$ on $\mathcal{T}=\left\{1,2,3,...,10\right\}$ and symmetric EV users are considered. As for the physical objective, and unless otherwise specified, simulations will be done considering Joule losses, that is $f^{Grid}(L_{t})=L_{t}^2$.

\medskip

\begin{figure}[!htbp]
\scalebox{0.6}
{
\begin{tikzpicture}[y=0.1cm,x=0.5cm,font=\sffamily]
 	%axis
	\draw (0,0) -- coordinate (x axis mid) (20,0);
    	\draw (0,0) -- coordinate (y axis mid) (0,100);
    	%ticks
    	\foreach \x in {5,10,15,20}
     		\draw (\x,-1) -- (\x,1) node[anchor=north,yshift=-0.6cm] {\x};
    	\foreach \y in {0,20,40,...,100}
     	\draw (-0.1,\y) -- (0.1,\y) node[anchor=east,xshift=-0.15cm] {\y}; 
	%labels      
	\node[below=0.8cm] at (x axis mid) {\textit{\Large Number of EV}};
	\node[rotate=90,above=0.8cm] at (y axis mid) {\textit{\Large Proportion of Nash equilibria} (\%)};

  %title
 \node[above=10cm] at (x axis mid) {
  	\centering\vspace*{\fill}
		\begin{minipage}{13cm}
		\centering
		\textbf{\Large Proportion of Nash equilibria of the atomic charging} \\
		\textbf{\Large game for $1$ to $20$ EV}
		\end{minipage}
		\vspace*{\fill}
		};
  
	%plots
	\draw[green,line width=\SizePlot pt] plot[mark=triangle*,mark options={fill=green,scale=\SizeMark}] 
		file {PartNENC3.txt};
	\draw[blue,line width=\SizePlot pt] plot[mark=square*,mark options={fill=blue,scale=\SizeMark}] 
		file {PartNENC4.txt};
	\draw[red,line width=\SizePlot pt] plot[mark=*,mark options={fill=red,scale=\SizeMark}] 
		file {PartNENC5.txt};
		
 %grid
 \draw[line width=0.2mm,dashed] (0,0) grid [xstep=3cm,ystep=2cm] (20,1);
   
	%legend
	%\fill [black!\CouleurFondLeg] (13.8,1.1) rectangle (20,1.2);
	\begin{scope}[shift={(7cm,8cm)}]
	\draw[red,line width=\SizePlot pt] (-0.5,0) -- 
		plot[mark=*, mark options={fill=red,scale=\SizeMark}] (0.25,0) -- (1,0) 
		node[right]{\large $C_{i}=C=5$};
	\draw[blue,line width=\SizePlot pt,yshift=2\baselineskip] (-0.5,0) -- 
	plot[mark=square*, mark options={fill=blue,scale=\SizeMark}] (0.25,0) -- (1,0)
	node[right]{\large $C_{i}=C=4$};
	\draw[green,line width=\SizePlot pt,yshift=4\baselineskip] (-0.5,0) -- 
	plot[mark=triangle*, mark options={fill=green,scale=\SizeMark}] (0.25,0) -- (1,0)
	node[right]{\large $C_{i}=C=3$};
	\end{scope}
\end{tikzpicture}
}
\caption{The proportion of Nash equilibria is combinatorial and globally decreasing with the number of EV.}
\label{FigPartNE}
\end{figure}
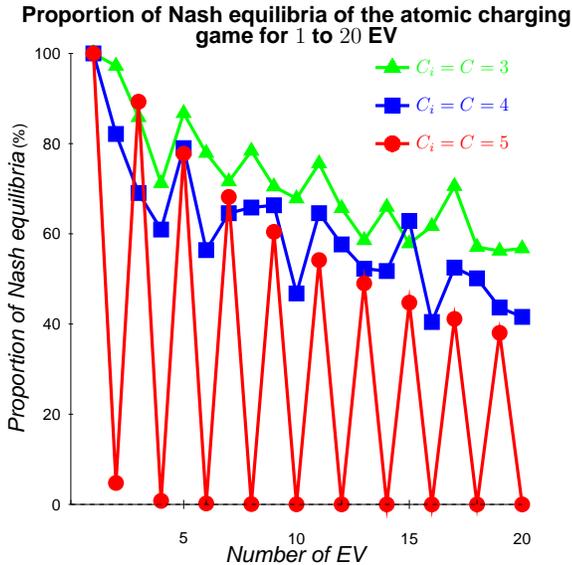

To begin with, we highlight that the proportion of Nash equilibrium is highly increasing with $I$ and also strongly depends on combinatorial effects between the charging duration $C$ and $T$ (see Fig.\ref{FigPartNE}).

%Proportion of NE = f(I)

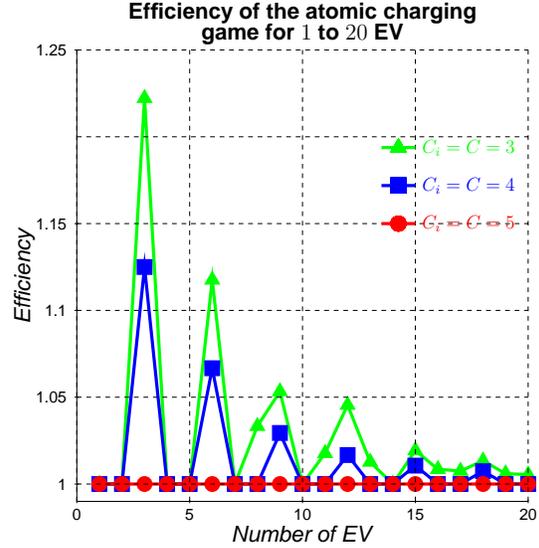
\begin{figure}[!htbp]
\scalebox{0.6}
{
\begin{tikzpicture}[y=38.5cm,x=0.5cm,font=\sffamily]
 	%axis
	\draw (0,0.99) -- coordinate (x axis mid) (20,0.99);
    	\draw (0,0.99) -- coordinate (y axis mid) (0,1.25);
    	%ticks
    	\foreach \x in {0,5,...,20}
     		\draw (\x,0.991) -- (\x,0.989) node[anchor=north] {\x};
    	\foreach \y in {1,1.05,1.1,1.15,1.25}
     	\draw (-0.1,\y) -- (0.1,\y) node[anchor=east,xshift=-0.15cm] {\y}; 
	%labels      
	\node[below=0.4cm] at (x axis mid) {\textit{\Large Number of EV}};
	\node[rotate=90,above=0.8cm] at (y axis mid) {\textit{\Large Efficiency}};

  %title
  \node[above=10cm] at (x axis mid) {
  	\centering\vspace*{\fill}
		\begin{minipage}{13cm}
		\centering
		\textbf{\Large Efficiency of the atomic charging} \\
		\textbf{\Large game for $1$ to $20$ EV}
		\end{minipage}
		\vspace*{\fill}
		};
  
	%plots
	\draw[green,line width=\SizePlot pt] plot[mark=triangle*,mark options={fill=green,scale=\SizeMark}] 
		file {PoANC3.txt};
	\draw[blue,line width=\SizePlot pt] plot[mark=square*,mark options={fill=blue,scale=\SizeMark}] 
		file {PoANC4.txt};
	\draw[red,line width=\SizePlot pt] plot[mark=*,mark options={fill=red,scale=\SizeMark}] 
		file {PoANC5.txt};
		
 %grid
 \draw[line width=0.2mm,dashed] (0,0.99) grid [xstep=2.5cm,ystep=1.925cm] (20,1.25);
    
	%legend
	%\fill [black!\CouleurFondLeg] (13.8,1.1) rectangle (20,1.2);
	\begin{scope}[shift={(14,1.15)}]
	\draw[red,line width=\SizePlot pt] (-0.5,0) -- 
		plot[mark=*, mark options={fill=red,scale=\SizeMark}] (0.25,0) -- (1,0) 
	node[right]{\large $C_{i}=C=5$};
	\draw[blue,line width=\SizePlot pt,yshift=2\baselineskip] (-0.5,0) -- 
	plot[mark=square*, mark options={fill=blue,scale=\SizeMark}] (0.25,0) -- (1,0)
	node[right]{\large $C_{i}=C=4$};
	\draw[green,line width=\SizePlot pt,yshift=4\baselineskip] (-0.5,0) -- 
	plot[mark=triangle*, mark options={fill=green,scale=\SizeMark}] (0.25,0) -- (1,0)
	node[right]{\large $C_{i}=C=3$};
	\end{scope}
\end{tikzpicture}
}
\caption{Efficiency of the atomic charging game for $C=3-5$ and a number of EV varying from $I=1$ to $20$}
\label{FigPoAN}
\end{figure}

\begin{figure}[!htbp]
\scalebox{0.6}
{
\begin{tikzpicture}[y=100cm,x=1.11cm,font=\sffamily]
 	%axis
	\draw (1,1) -- coordinate (x axis mid) (10,1);
    	\draw (1,1) -- coordinate (y axis mid) (1,1.1);
    	%ticks
    	\foreach \x in {1,2,...,10}
     		\draw (\x,0.999) -- (\x,1.001) node[anchor=north,yshift=-0.25cm] {\x};
    	\foreach \y in {1,1.02,1.04,1.06,1.08,1.1}
     	\draw (0.9,\y) -- (1.1,\y) node[anchor=east,xshift=-0.15cm] {\y}; 
	%labels      
	\node[below=0.7cm] at (x axis mid) {\textit{\Large Charging duration $C$}};
	\node[rotate=90,above=0.8cm] at (y axis mid) {\textit{\Large Efficiency}};

  %title
  \node[above=10.5cm] at (x axis mid) {
  	\centering\vspace*{\fill}
		\begin{minipage}{13cm}
		\centering
		\textbf{\Large Efficiency of the atomic charging game} \\
		\textbf{\Large for a charging duration from $C=1$ to $10$}
		\end{minipage}
		\vspace*{\fill}
		};
  
	%plots
	\draw[green,line width=\SizePlot pt] plot[mark=triangle*,mark options={fill=green,scale=\SizeMark}] 
		file {PoACN8.txt};
	\draw[purple,line width=\SizePlot pt] plot[mark=square*,mark options={fill=white,scale=\SizeMark}] 
		file {PoACN9.txt};
	\draw[blue,line width=\SizePlot pt] plot[mark=square*,mark options={fill=blue,scale=\SizeMark}] 
		file {PoACN10.txt};
	\draw[brown,line width=\SizePlot pt] plot[mark=triangle*,mark options={fill=white,scale=\SizeMark}] 
		file {PoACN11.txt};
	\draw[red,line width=\SizePlot pt] plot[mark=*,mark options={fill=red,scale=\SizeMark}] 
		file {PoACN12.txt};
		
 %grid
 \draw[line width=0.2mm,dashed] (1,1) grid [xstep=1.11cm,ystep=2cm] (10,1.1);
    
	%legend
	%\fill [black!\CouleurFondLeg] (13.8,1.1) rectangle (20,1.2);
	\begin{scope}[shift={(7,1.04)}]
	\draw[red,line width=\SizePlot pt] (-0.5,0) -- 
		plot[mark=*, mark options={fill=red,scale=\SizeMark}] (0.25,0) -- (1,0) 
	node[right]{\large $I=12$};
	\draw[brown,line width=\SizePlot pt,yshift=2\baselineskip] (-0.5,0) -- 
	plot[mark=triangle*, mark options={fill=white,scale=\SizeMark}] (0.25,0) -- (1,0)
	node[right]{\large $I=11$};
	\draw[blue,line width=\SizePlot pt,yshift=4\baselineskip] (-0.5,0) -- 
	plot[mark=square*, mark options={fill=blue,scale=\SizeMark}] (0.25,0) -- (1,0)
	node[right]{\large $I=10$};
	\draw[purple,line width=\SizePlot pt,yshift=6\baselineskip] (-0.5,0) -- 
	plot[mark=square*, mark options={fill=white,scale=\SizeMark}] (0.25,0) -- (1,0)
	node[right]{\large $I=9$};
	\draw[green,line width=\SizePlot pt,yshift=8\baselineskip] (-0.5,0) -- 
	plot[mark=triangle*, mark options={fill=green,scale=\SizeMark}] (0.25,0) -- (1,0)
	node[right]{\large $I=8$};
	\end{scope}
\end{tikzpicture}
}
\caption{Efficiency of the atomic charging game for $I=8-12$ and a charging duration from $C=1$ to $C=10$}
\label{FigPoAC}
\end{figure}
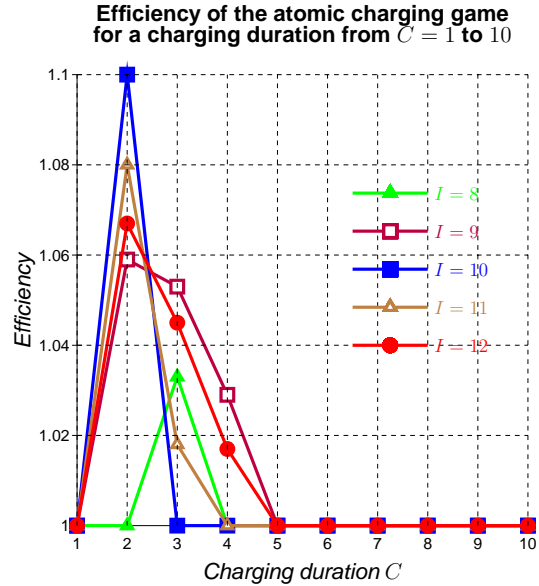

Simulations also provide a justification of the nonatomic charging game as the limiting case of the atomic formulation. On Fig.\ref{FigPoAN} it can be observed that, in spite of the cycles resulting from combinatorial effects between $T$ and $C$, the atomic efficiency is interestingly globally decreasing with the number of EV, approaching the nonatomic bound of one for the bigger values of $I$. Fig.\ref{FigPoAC} is the counterpart of the preceding showing how the atomic PoA approaches the limiting value of one when the number of EV is fixed, but the charging duration increases which implies that the global charging need also increases.

%PoA atomic = f(N)

%PoA atomic = f(C)

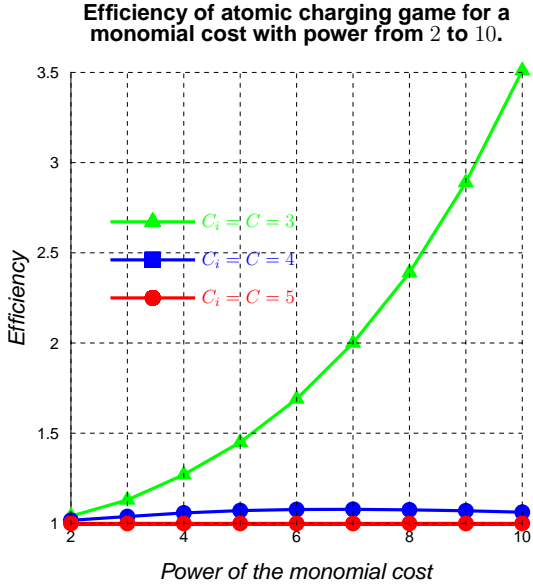
\begin{figure}[!htbp]
\scalebox{0.6}
{
\begin{tikzpicture}[y=4cm,x=1.25cm,font=\sffamily]
 	%axis
	\draw (2,0.99) -> coordinate (x axis mid) (10,0.99);
    	\draw (2,0.99) -> coordinate (y axis mid) (2,3.5);
    	%ticks
    	\foreach \x in {2,4,...,10}
     		\draw (\x,0.991) -- (\x,0.989) node[anchor=north] {\x};
    	\foreach \y in {1,1.5,2,2.5,3,3.5}
     	\draw (1.99,\y) -- (2.01,\y) node[anchor=east,xshift=-0.15cm] {\y}; 
	%labels      
	\node[below=0.7cm] at (x axis mid) {\textit{\Large Power of the monomial cost}};
	\node[rotate=90,above=0.8cm] at (y axis mid) {\textit{\Large Efficiency}};

  %title
  \node[above=10.5cm] at (x axis mid) {
  	\centering\vspace*{\fill}
		\begin{minipage}{13cm}
		\centering
		\textbf{\Large Efficiency of atomic charging game for a} \\
		\textbf{\Large monomial cost with power from $2$ to $10$.}
		\end{minipage}
		\vspace*{\fill}
		};
  
	%plots
	\draw[green,line width=\SizePlot pt] plot[mark=triangle*,mark options={fill=green,scale=\SizeMark}] 
		file {PoADegC3.txt};
	\draw[blue,line width=\SizePlot pt] plot[mark=*,mark options={fill=blue,scale=\SizeMark}] 
		file {PoADegC4.txt};
	\draw[red,line width=\SizePlot pt] plot[mark=*,mark options={fill=red,scale=\SizeMark}] 
		file {PoADegC5.txt};
		
 %grid
 \draw[line width=0.2mm,dashed] (2,0.99) grid [xstep=1.25cm,ystep=2cm] (10,3.5);
    
	%legend
	%\fill [black!\CouleurFondLeg] (13.8,1.1) rectangle (20,1.2);
	\begin{scope}[shift={(4cm,9cm)}]
	\draw[red,line width=\SizePlot pt] (-0.5,0) -- 
		plot[mark=*, mark options={fill=red,scale=\SizeMark}] (0.25,0) -- (1,0) 
		node[right]{\large $C_{i}=C=5$};
	\draw[blue,line width=\SizePlot pt,yshift=2\baselineskip] (-0.5,0) -- 
	plot[mark=square*, mark options={fill=blue,scale=\SizeMark}] (0.25,0) -- (1,0)
	node[right]{\large $C_{i}=C=4$};
	\draw[green,line width=\SizePlot pt,yshift=4\baselineskip] (-0.5,0) -- 
	plot[mark=triangle*, mark options={fill=green,scale=\SizeMark}] (0.25,0) -- (1,0)
	node[right]{\large $C_{i}=C=3$};
	\end{scope}
\end{tikzpicture}
}
\caption{Efficiency of the atomic charging game for a monomial cost $f^{Grid}=X^{k}$ with $k=2$ to $10$ and $C=3-5$.}
\label{FigPoADeg}
\end{figure}

Finally, Fig.\ref{FigPoADeg} presents the dependancy of the efficiency for monomial cost functions of the form $f^{Grid}(L_{t})=L_{t}^k$, showing, interestingly and contrary to standard results in the class of congestion games, that its value is not always increasing with the power $k$.

%%%%%%%%%%%%%%%%%%%%%%%%
\section{Conclusion}
%%%%%%%%%%%%%%%%%%%%%%%%

In this paper, an EV charging problem has been introduced and analyzed from a game theoretical standpoint to demonstrate some of its fundamental properties. Both atomic and nonatomic settings have been considered because of the different EV networks where this model may apply but also to investigate in more details the stronger properties of the nonatomic case as a limiting framework of the atomic problem.

\noindent

Proving that the proposed game belongs to the class of potential games, existence of a Nash equilibrium is obtained both when considering atomic or nonatomic users, while uniqueness is only valid in the nonatomic case. Showing that the unique Nash equilibrium coincides with the optimum on the electrical network side in the nonatomic case, this ensures that the "efficiency", a notion close to the standard Price of Anarchy, is one. In the atomic case, the problem is combinatorial and finding a tight bound is still an open problem. Simulations highlight some trends according to the values of parameters. In particular, the atomic "efficiency" get close to the nonatomic bound of one when the number of EV increases.

An analytical proof of this convergence may constitute an extension of this work. The physical modeling could also be enhanced introducing cost functions depending on the whole past of total load or allowing fractioned charging profiles using a battery aging model. Benefits on the network-side will indeed be a priori greater because the charging profiles will belong to a bigger set. Nonetheless, a coherent analysis will be needed to quantify the potential loss in terms of battery life.

%%%%%%%%%%%%%%%%%%%%%%%
\appendices

%%%%%%%%%%%%%%%%%%%%%%%%%%%%%%%%%%%%%%%%%%%%%%%%%
\section{Proof of the uniqueness of the Nash equilibrium in the first class of grid functions}
%%%%%%%%%%%%%%%%%%%%%%%%%%%%%%%%%%%%%%%%%%%%%%%%%
\label{ProofUniqNEClass1}

\begin{IEEEproof}
Take $f^{Grid}$ verifying Ass.\ref{H1}. Because $f^{Grid}$ is strictly increasing, the potential of the game $\Phi^{NA}$ is strictly concave. This ensures that $\Phi^{NA}$ has a unique maximum over $\Delta_{T-C}$, denoted by $\tilde{x}^{\star}$. To show that $\tilde{x}^{\star}$ is independent of $f^{Grid}$, we will demonstrate that, whatever $f^{Grid}$, the first order condition system of 

\begin{equation}
\label{MaxPot}
\max_{\tilde{x} \in \Delta_{T-C+1}} \Phi^{NA}(\tilde{x}) \text{,}
\end{equation}
has a unique positive solution, which will be $\tilde{x}^{\star}$ because of the potential property of the charging game.

\medskip

Consider the first order conditions 

\begin{equation}
\forall t, 1 \leq t \leq T-C, \, \frac{\partial \Phi}{\partial \tilde{x}^{\star}_{t}}=0 \textrm{.}
\end{equation}  

Particularizing $\tilde{x}_{T-C}$, given that 
\begin{equation}
\displaystyle \sum_{t=1}^{T-C} \tilde{x}_{t}=1 \textrm{,}
\end{equation}  
we have for $t<C$

\begin{eqnarray}
\frac{\partial \Phi}{\partial \tilde{x}_{t}}(\tilde{x}^{\star}) & = & \displaystyle \sum_{l=t}^{C} f^{Grid}(\displaystyle \sum_{u=1}^{l}\tilde{x}_{u}^{\star})+\displaystyle \sum_{l^{\prime}=1}^{t-1} f^{Grid}(\displaystyle \sum_{c=1}^{C}\tilde{x}_{l^{\prime}+c}^{\star}) \nonumber \\
& - & \frac{\partial \Phi}{\partial \tilde{x}_{T-C}}(\tilde{x}^{\star}) \\
& = & 0 \nonumber
\end{eqnarray} 
then for $C \leq t \leq T-2C+1$
\begin{eqnarray}
\frac{\partial \Phi}{\partial \tilde{x}_{t}}(\tilde{x}^{\star}) & = & \displaystyle \sum_{l=1}^{C} f^{Grid}(\displaystyle \sum_{c=1}^{C}\tilde{x}_{t+l-c}^{\star}) \nonumber \\
& - & \frac{\partial \Phi}{\partial \tilde{x}_{T-C}}(\tilde{x}^{\star}) \\
& = & 0 \nonumber
\end{eqnarray}
and finally for $T-2C+1 < t < T-C$
\begin{eqnarray}
\frac{\partial \Phi}{\partial \tilde{x}_{t}}(\tilde{x}^{\star}) & = & \displaystyle \sum_{l=1}^{T-C+1-t} f^{Grid}(\displaystyle \sum_{c=1}^{C}\tilde{x}_{t+l-c}^{\star}) \nonumber \\ 
& + & \displaystyle \sum_{l^{\prime}=1}^{t-T+2C-1} f^{Grid}(\displaystyle \sum_{u=1}^{l^{\prime}+1}\tilde{x}_{T-C+1-u}^{\star})-\frac{\partial \Phi}{\partial \tilde{x}_{T-C}}(\tilde{x}^{\star}) \nonumber \\
& = & 0
\end{eqnarray}

\medskip

This three different formulations correspond respectively to the first time intervals where the number of argument of $f^{Grid}$ is increasing, the central intervals where all arguments are sums of $C$ elements and the last where the sum's sizes decrease.

Substracting the second equation to the first and using the strict monotony of $f^{Grid}$, we have
\begin{equation}
\tilde{x}_{1}^{\star} = \displaystyle \sum_{t=2}^{C+1} \tilde{x}_{t}^{\star}
\end{equation}  
the third to the second
\begin{equation}
\tilde{x}_{1}^{\star}+\tilde{x}_{2}^{\star} = \displaystyle \sum_{t=3}^{C+2} \tilde{x}_{t}^{\star}
\end{equation}
and so on up to the substraction of the last equation ($t=T-C-1$) to the last but one ($t=T-C-2$). Keeping the last equation, which is 

\begin{eqnarray}
f^{Grid}(\displaystyle \sum_{c=1}^{C} \tilde{x}_{T-C-c}^{\star})-f^{Grid}(\tilde{x}_{T-C}^{\star})=0 \\
\Leftrightarrow \displaystyle \sum_{c=1}^{C} \tilde{x}_{T-C-c}^{\star}=\tilde{x}_{T-C}^{\star} \nonumber
\end{eqnarray}

and given that $\tilde{x}^{\star}$ sums to one, this gives a linear system of $T-C$ equations, independent of $f^{Grid}$. The matrix of this system, written here for $C=3$, is

\medskip

\begin{equation}
\label{MatLinSyst}
\begin{pmatrix}
1   & -1  & -1  & -1  & 0   & ... & 0   & 0   & 0   \\
1   & 1   & -1  & -1  & -1  &  0  & 0   & 0   & 0   \\ 
1   & 1   & 1   & -1  & -1  & -1  & 0   & 0   & 0   \\
0   & 1   & 1   & 1   & -1  & -1  &  -1   & ... & 0   \\
0   & 0   & 1   & 1   & 1   & -1  & -1  & -1   & ... \\
... & ... & ... & ... & ... & ... & ... & ... & ...  \\
... & ... & 0   & 1   & 1   & 1   & -1  & -1  & -1  \\
... & ... & ... & ... & 1   & 1   & 1   & -1  & -1  \\
... & ... & ... & ... & ... & 1   & 1   & 1   & -1  \\
1   &  1  & 1   &  1  &  1  &  1  &  1  &  1  &  1  \\
\end{pmatrix}
\end{equation}

This system is now demonstrated to have a unique and positive solution. Substracting the first line to the $C-1$ following lines with a $1$ on the first column, then repeating this transformation with the second and all the following lines to the last but one yields a matrix of the form

\begin{equation}
\label{MatLinSystBis}
\begin{pmatrix}
m_{1}   & -1   & -1     & -1     & 0      & ...    & 0   \\
0       & m_{2}& u_{2,3}& u_{2,4}& u_{2,5}&  ...   & u_{2,T-C+1}   \\ 
0       & 0    & m_{3}  & u_{3,4}& u_{3,5}& ...    & u_{3,T-C+1}    \\
0       & 0    & 0      & m_{4}  & u_{4,5}& ...    & u_{4,T-C+1}   \\
0       & 0    & 0      & 0      & m_{5}  & u_{5,6}& ...    \\
...     & ...  & ...    & ...    & ...    & m_{T-C}& u_{T-C,T-C+1}   \\
...     & ...  & 0      & 0      & 0      & 0      & m_{T-C+1} \\
\end{pmatrix}
\end{equation}

with all $m_{t}$ strictly positive because resulting of multiple substractions of $-1$ or $0$ above a $1$ and with all $u_{i,j}$ negative (substracting $-1$ multiplied by weights inferior to $1$ to $-1$). The second member of this equation, being initially

\begin{equation}
\label{SecMember}
\bm{b^{0}}=
\begin{pmatrix}
L_{C+1}^{exo}-L_{1}^{exo} \\
L_{C+2}^{exo}-L_{2}^{exo} \\
... \\
L_{T-1}^{exo}-L_{T-C-1}^{exo} \\
1
\end{pmatrix}
\end{equation}

is finally proven to be positive after all the transformations, which are enumerated with an index $\textrm{iter}$, the last being denoted by $\textrm{Iter}$. The second member after $\textrm{iter}$ operations is written $\bm{b^{\textrm{iter}}}$. Observe first that because $\bm{L^{exo}}$ is increasing, $b_{1}^{\textrm{iter}}=L_{C+1}^{exo}-L_{1}^{exo}$, which will not be modified by these transformations, will be always positive. The successive transformations on the lines consisting in substracting a line to its $C-1$ directly following lines and to the last, it can be expressed as
\begin{equation}
\label{PondInf1}
\left\{
\begin{array}{l}
Line_{t+1} \rightarrow Line_{t+1}-r^{iter}_{1}Line_{t} \\
Line_{t+2} \rightarrow Line_{t+2}-r^{iter}_{2}Line_{t} \\
...\\
Line_{t+C-1} \rightarrow Line_{t+C-1}-r^{iter}_{C-1}Line_{t} \\
Line_{T-C} \rightarrow Line_{T-C}-r^{iter}_{T-C}Line_{t} \\
\end{array}
\right.
\end{equation}

%\begin{equation}
%Line_{T-C+1} \rightarrow Line_{T-C+1}-r^{iter}_{T-C}Line_{t} \text{.}
%\end{equation}

The reader can then easily verify that 

\begin{equation}
\label{PondInf1Bis}
\forall iter, \, \forall c, \, 1 \leq c <C, r^{iter}_{c} \leq 1
\end{equation}

and that $r^{iter}_{T-C}=1$ for the $C$ first iterations, then $r^{iter}_{T-C}=2$ for the $C$ following and so on. Due to (\ref{PondInf1}-\ref{PondInf1Bis}), we have

\begin{equation}
\forall \textrm{iter}, \, \forall t, \, 1 \leq t<T-C, \, 0 \leq b^{\textrm{iter}}_{t} \leq L_{t+C}^{exo}-L_{t}^{exo}
\end{equation} 

which gives the positivity of the lines $2$ to $T-C-1$ of the second member $\bm{b}$ because

\begin{eqnarray}
\forall iter, \, b^{\textrm{iter}}_{t+c}-m_{c}b^{\textrm{iter}}_{t} & \geq & b^{\textrm{iter}}_{t+c}-b^{\textrm{iter}}_{t} \\
& \geq & \left[L_{t+2C}^{exo}-L_{t+C}^{exo}\right]-\left[L_{t+C}^{exo}-L_{t}^{exo}\right] \nonumber \\
& \geq & 0 \nonumber
\end{eqnarray}

using the convexity of $\bm{L^{exo}}$ in the last inequality. Finally, $b^{Iter}_{T-C}$ is given by
\begin{eqnarray}
b^{\textrm{Iter}}_{T-C} & = & 1-\displaystyle \sum_{k=1}^{q_{T,C}} (q_{T,C}-k+1) \left[L_{T-1-(k-1)C}^{exo}-L_{T-1-kC}^{exo}\right] \nonumber \\
& = & 1-q_{T,C}L_{T-1}^{exo}+\displaystyle \sum_{k=1}^{q_{T,C}}L_{T-1-kC}^{exo} \textrm{.}
\end{eqnarray}

Given assumption on $\bm{L^{exo}}$, $\bm{b}^{\textrm{Iter}}$ is thus positive. With the upper triangular structure of the modified matrix of the system and the positivity of $\bm{b}$, the positivity of $\bm{\tilde{x}}$ is obvious. Because this system is linear, this point is unique and is the unique maximum of $\Phi^{NA}$, which concludes the proof. 
\end{IEEEproof}

%%%%%%%%%%%%%%%%%%%%%%%%%%%%%%%%%
%\bibliographystyle{IEEEtran}
\bibliographystyle{IEEEtran_NoURL}
\bibliography{LittNETGCOP2}
\end{document}